\documentclass[lettersize,journal]{IEEEtran}
\usepackage{amsmath,amsfonts}
\usepackage{algorithmic}
\usepackage{algorithm}
\usepackage{array}
\usepackage[caption=false,font=normalsize,labelfont=sf,textfont=sf]{subfig}
\usepackage{textcomp}
\usepackage{stfloats}
\usepackage{url}
\usepackage{verbatim}
\usepackage{graphicx}
\usepackage{cite}
\usepackage{bm}

\begin{document}

\title{Analysis of Pleasantness Evoked by Various Airborne Ultrasound Tactile Stimuli Using Pairwise Comparisons and the Bradley-Terry Model}

\author{Sora SATAKE, Yoshihiro NAGANO, Masashi SUGIYAMA, Masahiro FUJIWARA, Yasutoshi MAKINO, \\and Hiroyuki
SHINODA
\thanks{This work was supported by JST CREST JPMJCR18A2. 
We would like to thank Editage (www.editage.com) for English language editing. 
\textit{(Corresponding author: Yasutoshi Makino.)} }
\thanks{S. Satake is with Graduate School of Information Science and Technology, The University of Tokyo, Tokyo, Japan. }
\thanks{Y. Nagano is with Graduate School of Informatics, Kyoto Univerisity, Kyoto, Japan. }
\thanks{M. Fujiwara is with Department of Electronics and Communication Technology, Nanzan University, Nagoya, Japan. }
\thanks{M. Sugiyama, Y. Makino, and H. Shinoda are with Graduate School of Frontier Sciences, The University of Tokyo, Chiba, Japan. }
}



\maketitle

\begin{abstract}
The presentation of a moving tactile stimulus to a person's forearm evokes a pleasant sensation. 
The speed, intensity, and contact area of the strokes should be systematically changed to evaluate the relationship between pleasantness and tactile stimuli in more detail. 
Studies have examined the relationship between stroking stimulation and pleasant sensations using airborne ultrasound tactile displays. The ultrasound-based method has the advantage of reproducible control of the speed, intensity, and contact area of the stimulus. 
In this study, we prepared new stimuli focusing on the modulation methods and the contact area and aimed to clarify their relationship with pleasantness in more detail.
Evaluating subjective sensations, such as pleasantness, numerically and consistently is challenging, warranting evaluation based on comparison. 
We propose a stimulus evaluation method that combines rough evaluation using Likert scales, detailed evaluation using pairwise comparisons, and quantification of comparison data using the Bradley--Terry model. 
As a result, we confirmed that the stimulus using lateral modulation and that with a large contact area used in this study were more pleasant than the conventional stimulus for six out of ten participants.
\end{abstract}


\section{Introduction}
\IEEEPARstart{T}{actile} sense has two roles. 
First, the discriminative role provides information about the external world, such as the shape, texture, and temperature of objects. 
Second, the affective role is crucial to social interactions and communication~\cite{DiscrimAndAffect}. 
In other words, people sometimes use bodily touch to communicate their emotions. 
For example, anger or love can be communicated using only tactile stimuli~\cite{emotion}.

Among emotional tactile stimuli, stroking on the forearm has attracted much attention. 
Stroking has been shown to evoke pleasant feelings~\cite{pleasant} and is often used to convey affection and sympathy~\cite{emotion}. 
It is known that stroking the arm with a brush at a speed of 10 - 100 mm/s maximizes the evoked pleasantness, and that the firing rate of nerve fibers called C-tactile fibers is correlated with the pleasantness. 

Studies have also replicated the effects of stroking using tactile devices. 
Some devices stroke the skin directly~\cite{StrokingDevice1,StrokingDevice2}, while others replicate  stroking through vibration, pressure, skin deformation, etc., utilizing the illusion of apparent motion~\cite{StrokingVib,StrokingPress,StrokingSlip,StrokingVPS}. 
Regarding vibrations, a smaller perceived intensity of the stimulus may be more pleasant~\cite{StrokingVib}, suggesting that factors other than stroke speed may affect pleasantness. 
If the factors are investigated so that more pleasant tactile stimuli can be designed, they will be even more useful in telecommunications and other applications. 
Therefore, the relationship between pleasantness and the nature of tactile stimuli should be discussed in more detail by systematically varying the speed, intensity, and contact area of stroking. 

In this study, we applied stroking stimuli to the forearm using airborne ultrasound tactile display (AUTD)~\cite{AUTD} and investigated the pleasantness of the stimuli. 
A previous study has confirmed that moving point stimuli using AUTD can evoke pleasant sensations~\cite{AUTD_ple}. 
In addition, AUTD is suitable for the investigation of pleasant tactile stimuli considering the stroking speed, perceived intensity, and area for several reasons. 
\begin{itemize}
    \item[1)] AUTD can present smoothly moving stimuli by updating their position at a frequency of 1 kHz. It can also freely change the speed of stroking by changing the time interval of changing the stimulus position. 
    \item[2)] Several modulation techniques, such as Amplitude Modulation (AM) and Lateral Modulation (LM), can be used to present moving stimuli with vibration at various perceived intensities. 
    \item[3)]AUTD can be applied to multiple points simultaneously, allowing the stimulus area to be easily changed. 
\end{itemize}
This study aimed to prepare new stimuli focusing on the modulation method and perceived area of the stimuli, clarify the relationship with pleasantness in more detail, and discover relationships that can make stimuli more pleasant. 

Evaluating subjective sensations such as pleasantness numerically and consistently is challenging. 
For example, evaluating pleasantness using a Likert scale may not provide consistent data because the evaluation scale varies from participant to participant or changes over time~\cite{compare}. 
In addition, if the experiment was conducted using a 7-level scale, the same evaluation value would be given to stimuli with slight differences in pleasantness.
Thus, participants may not be able to express detailed differences in their preferences. 
To address these issues, pairwise comparisons should be introduced because comparisons between two stimuli are much easier than absolute evaluations of the stimuli alone. 

In this study, we propose a new evaluation method that combines rough evaluation using a Likert scale with detailed evaluation using pairwise comparison. 
First, the participants were asked to rate each stimulus on a Likert scale. Stimuli with significantly different evaluation values were not compared further, while those with similar evaluation values were individually compared pairwise. 
Doing so reduced the number of comparisons and the burden on the participants, compared with those associated with comparing all stimuli with a reference stimulus~\cite{softness}. 
These individual relative evaluation values were quantified using the Bradley--Terry (BT) model and used as the final evaluation values. 

Dittrich et al.~\cite{Likert2BT} proposed a method to evaluate Likert scale data by converting them into pairwise comparison data and then reconstructing them using the BT model. 
In contrast, using our proposed method after obtaining an approximate trend using a simple Likert scale, the data were refined by pairwise comparisons to clarify detailed relationships. 
The proposed method is novel in that the overall structure of the rough evaluation can be refined by repeated pairwise comparisons for targets that are difficult to evaluate in absolute terms, such as pleasantness. 

The two main contributions of this study are as follows: 
\begin{enumerate}
    \item The design of more pleasant stimuli focusing on perceived intensity, area, and vibration.
    \item The evaluation of pleasantness by combining rough evaluation using a Likert scale and detailed evaluation using pairwise comparisons.
\end{enumerate}

\section{Experimental system}
\subsection{Apparatus}
In this study, we presented a moving stimulus that imitates stroking using AUTD~\cite{AUTD}. 
AUTD is an ultrasound--phased array consisting of 249 transducers per unit. The phase and amplitude of the ultrasound emitted from each transducer are controlled to form a focus in the air, and the tactile sensation was presented by the acoustic radiation pressure. 

\begin{figure}[htbp]
 \centering
 \includegraphics[width=2.0in]{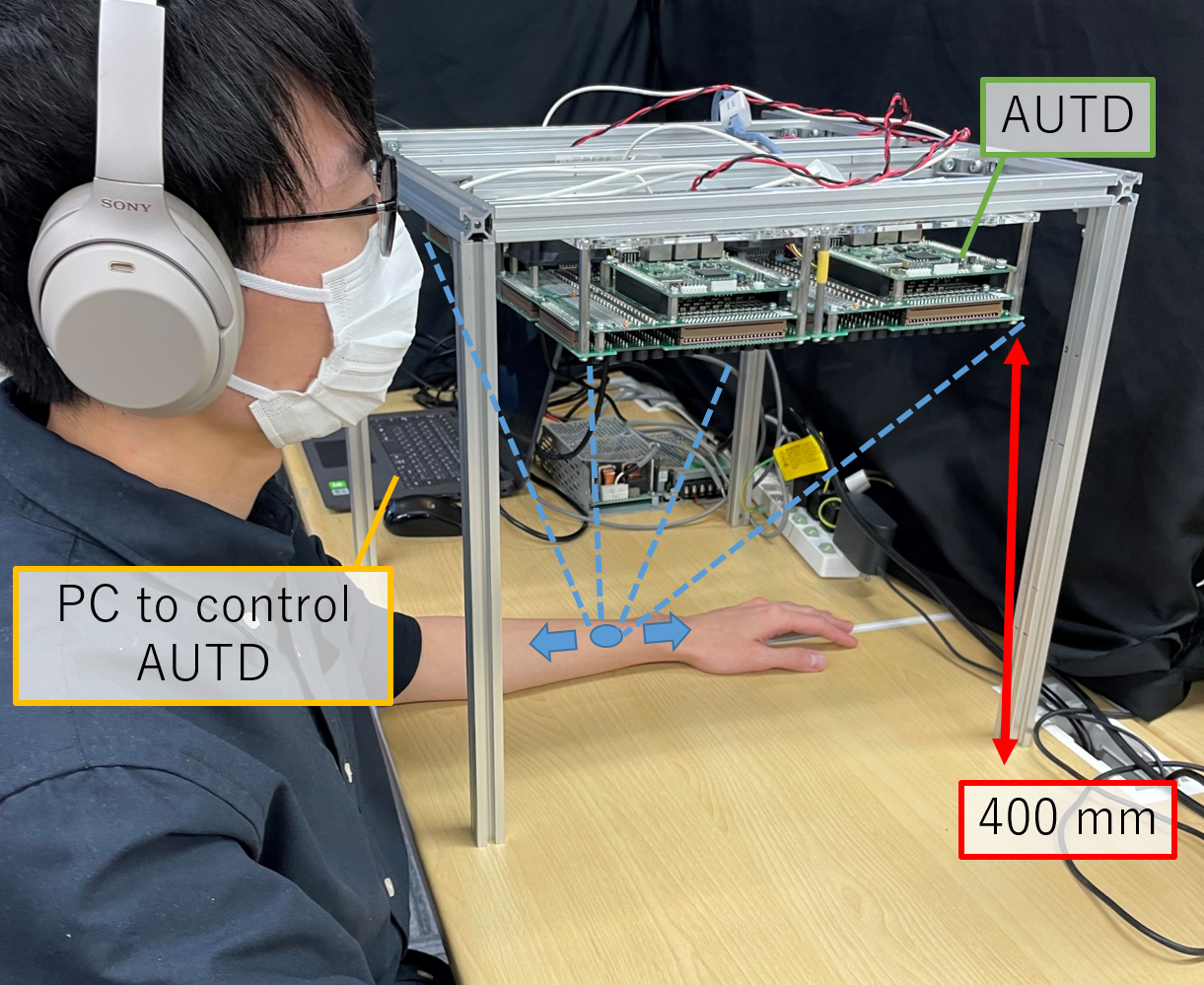}
 \caption{Experimental apparatus and scene of the experiment}
 \label{exp}
\end{figure}

\begin{figure}[htbp]
 \centering
 \includegraphics[width=2.0in]{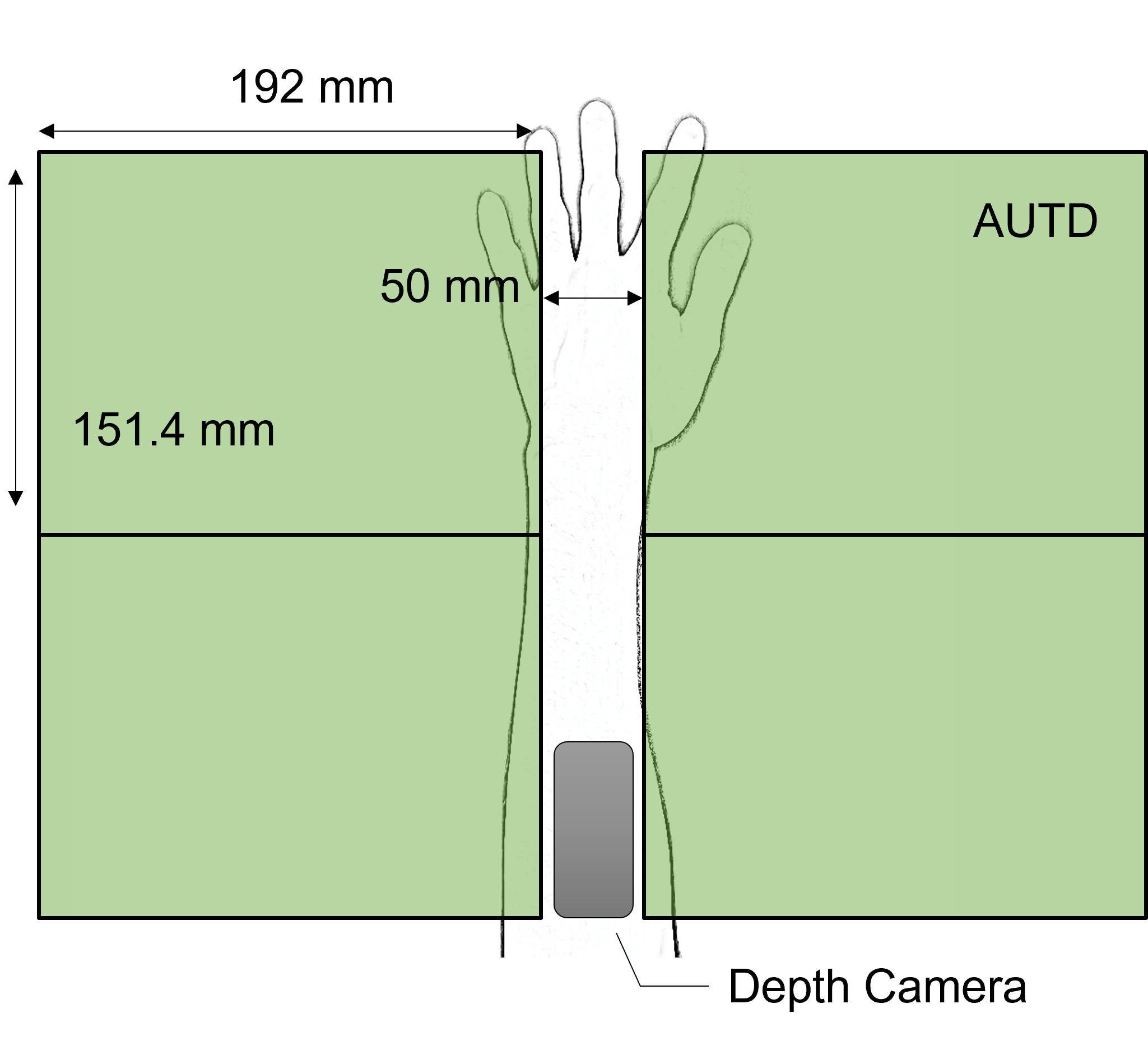}
 \caption{Top view of the AUTD arrangement}
 \label{haiti}
\end{figure}

The experimental apparatus is shown in Fig.\ref{exp} and Fig.\ref{haiti}.
Four AUTDs were placed approximately 400 mm away from the desk, facing downward. 
The participants placed their arms under the AUTDs and ultrasound stimuli were applied to the surface of the forearm.
A depth camera (Intel Realsense Depth Camera D435) was placed between the AUTD units to determine the position of the stimulus.

\subsection{Stimuli}
\subsection*{Stroking stimulus with AUTD}
The size of the ultrasonic focus of the AUTD is approximately equal to the wavelength; therefore, if 40 kHz is used, the diameter is approximately 10 mm.
Thus, if the focus is moved at intervals of 1 mm, which is sufficiently smaller than the diameter of the focal point, the tactile stimuli are felt to move continuously. 
In this study, we used this techniqe as the basic presentation method for stroking and prepared five types of stimulus patterns (static pressure, AMs, LMs (high/low frequency), and two-point stimuli) by changing the modulation scheme and number of foci. 

In addition, the speed of focal point movement was controlled by changing the time interval to update the focal point position. Three different speeds (50, 100, and 300 mm/s) were used in this study. 

The stimuli were presented to the posterior forearm of the left arm (the back side of the forearm), from near the elbow to 150 mm toward the wrist, as shown in Fig.~\ref{exp}. 
A depth camera was used to measure the three-dimensional shape of the surface of the presentation area, and the depth of focus was adjusted accordingly.
The stimulus presentation time for each trial was set to 3 s, regardless of the speed. 
The details of the five types of stimulus patterns are presented below.

\subsubsection{Static pressure and AM stimuli}
The static pressure stimulus was a stroking stimulus without modulation. The ultrasound focus of the constant sound pressure was moved at three different velocities. 
In contrast, with the AM stimulus the amplitude of the ultrasound focus was periodically modulated, providing a vibration sensation~\cite{AM}. 
In this study, we used 200 Hz sinusoidal modulation. 

These are the same stimuli used in a previous study\cite{AUTD_ple} and are intended for comparison with the results of that study. The following sections describe the new stimuli used in this study. 

\subsubsection{LM stimuli}
With the LM stimulus, the stimulus position is periodically changed, rather than the amplitude of the ultrasound, to increase the perceived intensity and provide a vibrating sensation\cite{LM}. 

\begin{figure}[htbp]
 \centering
 \includegraphics[width=2.0in]{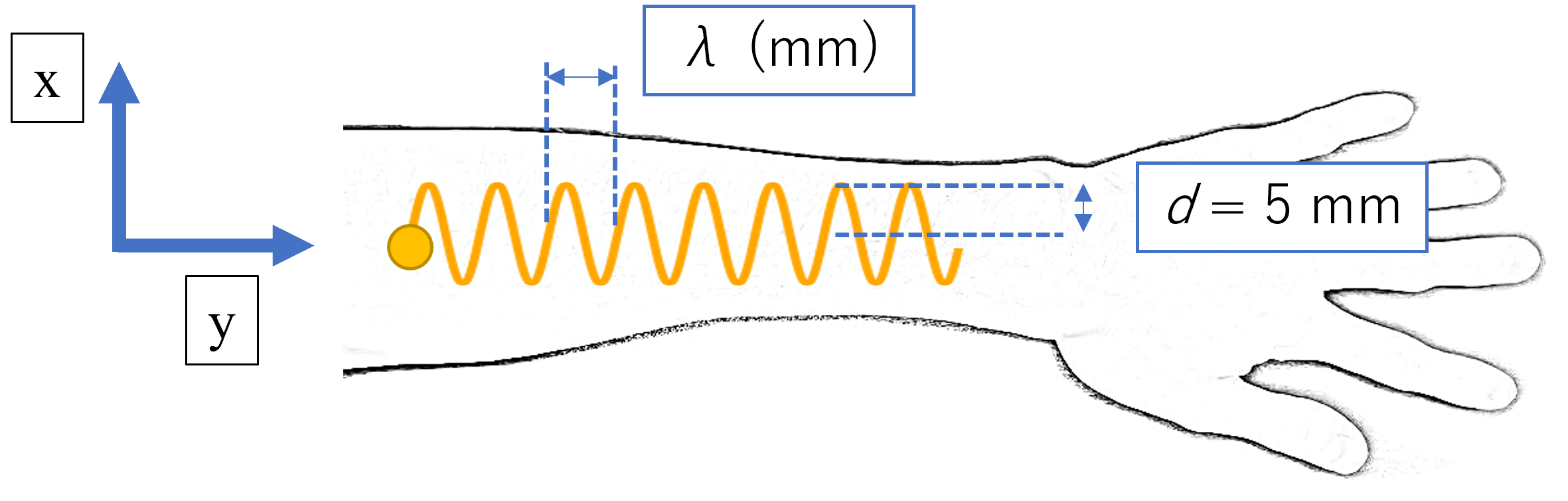}
 \caption{Schematic diagram of LM stimulation. $\lambda$ represents the spatial wavelength of the vibration (15 mm, 1.5 mm).}
 \label{LM}
\end{figure}

The pattern of the LM stimuli presented in this study is shown in Fig.~\ref{LM}. When the arm surface is considered to be an xy--plane, a focal point moving at a constant velocity along the $y$-axis oscillating sinusoidally in the $x$-axis direction is presented. Two types of stimuli, low and high, were prepared according to spatial wavelength, $\lambda$. 
Under low stimulation, $\lambda = 15$ mm, whereas under high stimulation, $\lambda = 1.5$ mm. The displacement $d$ in the $x$ direction was kept constant at $\pm5$ mm, regardless of the conditions. The moving speeds were 50, 100, and 300 mm/s, which were the same as those under static conditions. 
The spatial vibration frequency of the point is given by:
$$f=\frac{v}{\lambda}.$$
For example, $f = \frac{100}{15}$ Hz under low conditions, at a travel speed of 100 mm/s.

\subsubsection{Two-point stimulus}
A two-point stimulus is a stimulus in which two adjacent focal points on the skin are designated. 
By setting the distance between the centers of the two foci to 5 mm, which is smaller than the diameter of the stimulus, we aimed to create the sensation of a moving stimulus with a large perceived area rather than two separate points.
The four AUTD units were divided into two sections on the left and right sides to create two independent focal points. By halving the number of units per focal point, the sound pressure decreased, while the stimulus area increased.  
In addition, the conditions were identical to those of static pressure stimulation.

\section{Combination of absolute evaluation and pairwise comparisons}
\label{section_3}
\subsection{Pairwise comparisons between stimuli}
\label{section_3_1}

\begin{itemize}
    \item[1)] Roughly evaluate pleasantness of stimuli using a 7-level Likert scale.
    \item[2)] Obtain detailed comparison data by pairwise comparison of stimulus pairs with similar ratings.
    \item[3)] Create a comparison data-set based on the evaluation in step 1),  and add the results of step 2).
    \item[4)] Convert back to absolute value evaluation from the comparison data-set using the BT model, and normalize. 
\end{itemize}
The procedure is described in detail in the following sections. 

First, the participants were asked to rate the pleasantness of the stimuli on a 7-level scale in the ``pre-evaluation.'' 
This evaluation had the same issues as the Likert scale described above. 
Therefore, we conducted pairwise comparisons between two stimuli and evaluated the pleasantness of the stimuli in detail. 

Based on the results of the pre-evaluation, the accuracy of the evaluation results was improved by repeating detailed pairwise comparisons with stimuli that were considered to have similar levels of pleasantness. 
Specifically, we focused on the comparison between stimuli rated within two points to analyze detailed preferences while omitting the comparison of apparently different stimuli with a rating difference of more than three levels. 
The detailed procedure is described in Section \ref{method}.

Such pairwise comparisons provide individual comparative data; for example, $A$ is always rated more pleasant than $B$. 
For omitted comparison pairs, the results of the pre-evaluation were converted to the comparison results. 
For example, for stimulus $A$ with a pre-evaluation value of +2 and stimulus $B$ with a pre-evaluation value of -3, comparison data were added assuming that $A$ was evaluated as more pleasant than $B$ because the outcome was thought to be obvious. 
This method was expected to preserve the pre-evaluation values as the base data to some extent. 
In summary, for pairs that were actually compared, the results were adopted, and for pairs that omitted comparison, the pre-evaluation results were converted to pair-wise comparison results. 
Thus, a comparative data-set among all stimuli is obtained. 
The proposed method is novel in that it integrates the results of the pre-evaluation and pairwise comparison to estimate from the BT model. 
In the next section, this comparative data-set is converted to quantitative values based on the BT model.

\subsection{BT model}
Repeating pairwise comparisons between $n$ different stimuli to determine which stimulus is pleasant yields the probability of judging the $i$th stimulus as pleasant in the $i$th and $j$th stimulus pair. 
In the BT model, this probability was assumed to follow the following equation:
$$
P_{ij}=\frac{\pi_i}{\pi_i+\pi_j},
$$
where $\pi_i$ corresponds to the intensity of the $i$th stimulus and thus, represents the degree of pleasantness of the stimulus. 

In this model, unknown parameters are defined as ${\bm \pi} = \{\pi_1, \pi_2, ..., \pi_n\} $. We performed maximum likelihood estimation to calculate the optimal parameters from the comparison data-set using the Iterative Luce Spectral Ranking algorithm~\cite{choix}. 
For the implementation, we used choix, a Python library that provides an inference algorithm for models based on Luce's choice axiom. 

We then normalized the estimated ${\bm \pi}$ for each participant. 
This resulted in the alignment of the scale for each participant, where ``$0=$ most unpleasant'' and ``$1=$ most pleasant.'' 
Furthermore, for all data, we multiplied by 6 and then subtracted 3. 
This operation results in a scale of ``$-3=$ most unpleasant'' and ``$+3=$ most pleasant,'' which corresponds to the values from the pre-evaluation. 
The pleasantness values obtained during this procedure were used for the analysis.

\subsection{Experimental procedure}
\label{method}
The experiment was conducted on ten participants (six males and four females) aged 20--29. 
The stimuli were presented in five different patterns (static, AM, two-point, LM low, and LM high) at three different movement speeds, resulting in fifteen different stimuli. 
During the experiment, the participants wore headphones and listened to white noise to eliminate auditory information. 
Visual information was not controlled. 
Also, we did not control for what participants imagine in their minds when they evaluated the pleasantness of stimuli.
All procedures were performed in accordance with the Declaration of Helsinki (2008) and the participants provided informed consent prior to experimental commencement. 

In the first step of the experiment, a pre-evaluation was conducted to evaluate the pleasantness of each stimulus. 
First, the participants were asked to experience all the stimuli individually assuming that they would evaluate the pleasantness of the stimuli. Participants were also instructed that the total number of stimuli was 15.

Next, the 15 stimuli were randomly divided into 5 groups of 3 stimuli each, and each group was presented to the participants. The participants then selected the most pleasant and unpleasant stimuli in each group.
Each of the five pleasant and unpleasant stimuli selected in this way were presented again, and the participants were asked to choose the most pleasant and unpleasant stimuli among the five. 
Using the selected stimuli as the $+3$ and $-3$ criteria, participants were asked to rate the pleasantness of the remaining 13 stimuli on a 7-level scale from $+3$ (pleasant) to $-3$ (unpleasant). This procedure completes the pre-evaluation process.

In the second stage of the experiment, pairwise comparisons were conducted based on the evaluation values of each stimulus. Based on the 7-level evaluation values obtained in the pre-evaluation, pairs with the same rating were compared twice, and pairs with a 1-point difference were compared once. 
Participants were asked to respond to the more pleasant stimulus in each pair. 
Averaged across all participants, 13.1 pairs were compared twice, 27.7 pairs were compared once, and 64.2 pairs were omitted. 
This resulted in 45--66 pairwise comparisons (average 53.9) to judge the pleasantness of the 15 stimuli.

Because comparing every pair once each would require 105 comparisons, our proposed method reduced the number of pairwise comparisons by about half and enabled us to examine pairs that we wanted to investigate in more detail twice. 
The total time of the experiment was approximately 30 min.

\section{Analysis}
The values of pleasantness obtained by the method in Section~\ref{section_3} were used as the ``after-comparison'' data. 
In contrast, the pre-evaluation results were used as ``before-comparison'' data for the analysis. 

To examine the effectiveness of pairwise comparisons, we compared the data before and after the pairwise comparisons. 
The purpose of this method was to evaluate detailed preferences while preserving general preferences. 
Therefore, we first calculated the correlation coefficients between the data before and after the pairwise comparison for each participant and confirmed whether the general preferences were preserved. 
Next, we calculated the Mean Absolute Difference of the data before and after the pairwise comparison for each participant and quantitatively evaluated the change in the average values by pairwise comparison. 

To confirm whether more pleasant stimuli could be designed, the evaluation values of all the participants were averaged and compared. We observed large individual differences in the evaluation of pleasantness, and averaging the values would result in masking the differences in individual preferences; therefore, we also examined the after-comparison data for each participant individually. 

\begin{figure*}[t]
 \begin{center}
  \includegraphics[width=6.5in]{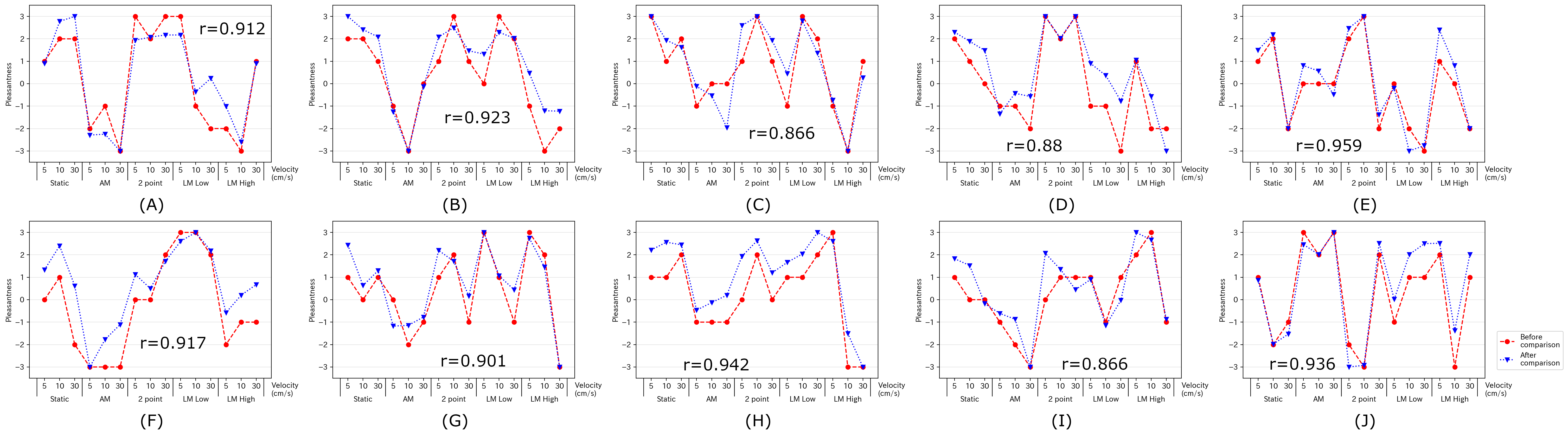}
  \caption{Results of before and after pairwise comparisons for each participant. The red dashed lines correspond to before comparison and the blue dotted correspond lines to after comparison. $r$ indicates the correlation coefficient between the before and after pairwise comparison data for each participant.}
  \label{all_subject}
 \end{center}
\end{figure*}

\section{Results}
\subsection{Effectiveness of pairwise comparisons}
Fig.~\ref{all_subject} shows the data of the pre-evaluation (shown in red) and evaluation after conducting pairwise comparisons (shown in blue) of each participant.
In the same figure, we also show the correlation coefficient $r$ between the before and after pairwise comparison data for each participant. 
The correlation coefficients were relatively large ($>$ 0.8) for all participants. 
In other words, the general form of the evaluation at the time of pre-evaluation was retained. 
Furthermore, the stimuli with the same evaluation values in the pre-evaluation are shown by the blue line, indicating that we succeeded in refining similar stimuli. 

\begin{figure}[htbp]
 \centering
 \includegraphics[width=2.0in]{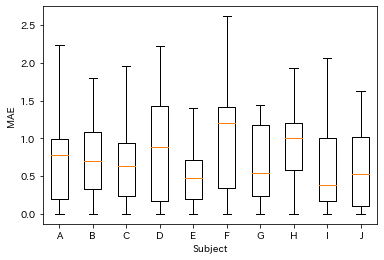}
 \caption{Mean absolute difference of pleasantness before and after the pairwise comparison for each participant. Whiskers indicate maximum and minimum values.}
 \label{fig_MAE}
\end{figure}

Fig.~\ref{fig_MAE} shows box plots of the mean absolute difference of the before and after pairwise comparison for each participant. 
The changes before and after the pairwise comparisons are concentrated at 0.5--1.0, and most of the changes are $<$ 2.0, with some exceptions.

\subsection{Designing more pleasant stimuli}
\begin{figure}[htbp]
 \centering
 \includegraphics[width=2.0in]{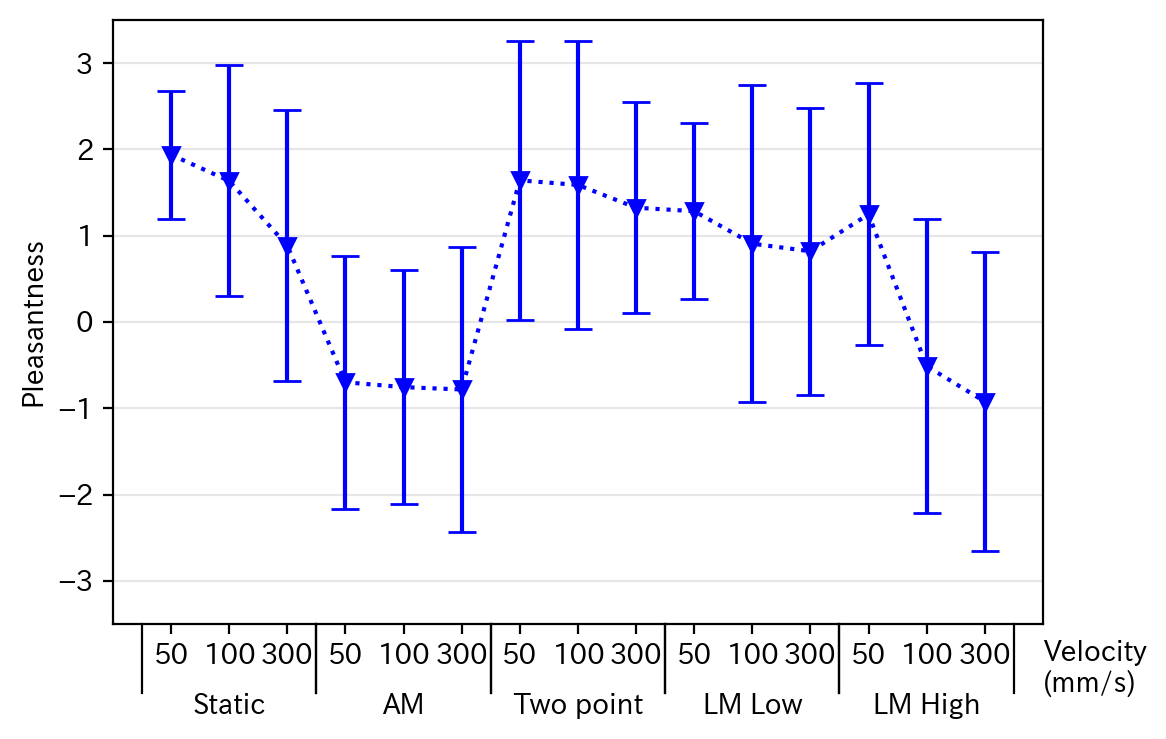}
 \caption{Mean pleasantness of all participants after pairwise comparisons. Error bars indicate standard deviation.}
 \label{fig_mean}
\end{figure}

Fig.~\ref{fig_mean} shows the means values of all participants for the results after pairwise comparisons. 
Error bars indicate standard deviation. The six conditions on the left are those used in previous studies, and the nine conditions on the right show the stimuli proposed in this study.
The two-point and LM low stimuli evoked a relatively pleasant sensation, similar to the previously proposed static stimulus. 
However, large individual differences were observed, and factors with high subjectivity, such as pleasantness, were difficult to evaluate using averages.

\section{Discussion}
\subsection{Effectiveness of pairwise comparisons}
From Fig.~\ref{all_subject}, the correlation coefficients before and after the pairwise comparisons are relatively large, suggesting that the preferences were roughly maintained before and after the comparisons. 
The general shape of the graph is similar for all participants, with some minor differences. 

Moreover, Fig.~\ref{fig_MAE} shows that most of the changes in the scores before and after the pairwise comparisons were $<$ 2.0, which could be because in determining the pairs for pairwise comparisons in the proposed method, comparisons were omitted if the values of the ratings differed by two or more. 

Among the changes that occurred before and after the comparison, ``differences'' and ``reversals'' are particularly noteworthy. 
The graph of participant (F) in Fig.~\ref{all_subject} shows that some stimulus groups that had the same ratings before the comparison shown in the red dots differed after the comparison.
For example, for AM speeds of 50, 100, and 300 mm/s, the pleasantness was -3 before the comparison. 
However, after the comparison, there was a difference in values, and the order among the stimuli became clear, as paired comparisons could express subtle differences that could not be determined by the Likert scale alone. 

In some cases, such as that of 100 and 300 mm/s two-point static pressure for participant (F), the evaluations were reversed before and after the comparison. 
This may be because pairwise comparisons enabled clear judgments without ambiguity, whereas judgments were ambiguous when they were evaluated using a Likert scale.
Based on the above, we confirmed that a combination of numerical absolute evaluations using Likert scales and pairwise comparisons can be used to evaluate detailed preference trends in a time-efficient manner.

This method also has the advantage of being less sensitive to the effects of previous stimulus histories.
Evaluating individual stimuli using a Likert scale does not allow participants to ignore the influence of the immediately preceding stimulus on how they feel and their psychological state at the time of the response.
For example, if similar stimuli continue to be presented, participants may become bored, and their pleasantness may decrease.
Moreover, in some cases, participants may expect a response other than +2 if only a +2 response has been previously recorded. 
Conventionally, these variations have been addressed by randomizing the order of presentation and averaging the responses by presenting the stimuli multiple times; however, this requires a large number of trials to completely eliminate variation. 
However, the proposed method reduces this effect by allowing participants to concentrate on comparing the two stimuli presented and eliminating the need to use past stimulus sequences as a criterion for judgment.

\subsection{Designing more pleasant stimuli}
Fig.~\ref{fig_mean} suggests that no new stimulus was more pleasant than a conventional stimulus (static pressure). 
In contrast, Fig. ~\ref{all_subject} shows that many participants seemed to prefer new stimuli to conventional stimuli. 
According to both the maximum and mean values and the preferences classified by the type of stimuli, participants (A)--(C) preferred static pressure, (D) and (E) preferred two--point pressure, (F)--(H) preferred LM low, (I) preferred LM high, and (J) preferred AM. 
In other words, participants (D)--(I) preferred the newly proposed stimuli. 
Therefore, we achieved our goal of designing stimuli that are more pleasant than conventional stimuli, despite the differences in individual preferences. 
We also found that analysis focusing on individual differences is necessary because some results may be missed by averaging. 

Some participants may have been strongly influenced not only by the type of stimulus but also by velocity. 
For example, participants (D), (E), (G), and (I) showed a relatively strong tendency for the evaluation of pleasantness to decrease as the speed of stroking increased.  
This is consistent with the results of a previous study\cite{pleasant}, which showed that a velocity of 10--100 mm/s is optimal for evoking pleasant sensations by stroking and that pleasantness decreases outside this range. 
Other trends, such as a U-shape and inverted U-shape  with respect to the velocity, have also been observed and require more detailed investigation. 

As previous studies have shown that the velocity that can evoke a pleasant sensation by stroking the arm is 10--100 mm/s, we adopted velocities of 50 and 100 mm/s, which are within the range, and 300 mm/s, which is outside the range. 
However, the optimal velocity in ultrasound stimulation is not necessarily 10--100 mm/s, as in the existing results, owing to the absence of frictional force. 
Therefore, further studies should investigate speeds below 50 mm/s. 

In the future, more pleasant stimuli common to all participants should be discovered while analyzing individual differences. 

\subsection{Optimal Number of Comparison}
This method is limited by its small number of comparisons. 
After the pre-evaluation, pairwise comparisons were performed twice for those with no point difference and once for those with a difference of two points or less. 
Consequently, the number of comparisons was reduced by about half. 
However, whether this is sufficient to detect detailed differences should be verified. 
In addition, pairs that were omitted because the comparison result was obvious may have yielded the opposite result when compared.
Therefore, a careful examination of the omitted conditions is also necessary. 
From the viewpoint of responding to changes in preferences over time, further improvement can be considered in the future, such as an experimental protocol in which comparisons are made at least once in the first half and once in the second half of the experiment, or a model that allows analysis over time.

\section{Conclusion}
In this study, we designed new pleasant tactile stimuli using AUTD. 
The combination of the Likert scale and pairwise comparisons enabled us to efficiently evaluate pleasantness in terms of time, reflecting the detailed preferences of the participants. 
Furthermore, using the BT model, the comparison data were converted into quantitative values for analysis. 
Consequently, we confirmed that pairwise comparison is an effective method for evaluating pleasantness, which is difficult to evaluate in absolute terms. 
In addition, we succeeded in designing tactile stimuli that were more pleasant than conventional stimuli, although preferences varied by individual. 
Future work will include the study of experimental design and analysis models that can cope with changes in preferences over time and the design of pleasant stimuli that consider individual differences. 
%


\bibliographystyle{IEEEtran}
\bibliography{references}

\end{document}